\documentclass[%
 reprint,
 amsmath,amssymb,
 aps,
]{revtex4-2}

\usepackage{graphicx}
\usepackage{xcolor} 
\usepackage{dcolumn}
\usepackage{bm}
\usepackage{ulem} 

\definecolor{myblue}{rgb}{0.05,0.1,0.5}

\definecolor{myred}{rgb}{0.5,0.05,0.1}

\begin{document}

\title{Caustic-like Structures in UHECR Flux after Propagation in Turbulent Intergalactic Magnetic Fields}
\author{K.~Dolgikh$^{1,2}$, A.~Korochkin$^{3,2}$,  G.~Rubtsov$^{1,2}$, D.~Semikoz$^{4}$, and I.~Tkachev$^{1,2}$}
\affiliation{$^{1}$ Institute for Nuclear Research of the Russian Academy of Sciences, Moscow, 117312 Russia}
\address{$^{2}$Physics Department and Laboratory of Cosmology and Elementary Particle Physics,Novosibirsk State University, Novosibirsk, 630090 Russia}
\affiliation{$^{3}$ Université Libre de Bruxelles, CP225 Boulevard du Triomphe, 1050 Brussels, Belgium}
\affiliation{$^{4}$ APC, Universit\'e Paris Cit\'e, CNRS/IN2P3, CEA/IRFU, Observatoire de Paris, 119 75205 Paris, France}

\begin{abstract}
UHECR propagation in a turbulent intergalactic magnetic field in the small-angle scattering regime is well understood for propagation distances much larger than the field coherence scale. The diffusion theory doesn't work and unexpected effects may appear for propagation over smaller distances, from a few and up to 10-20 coherence scales. We study the propagation of UHECRs in this regime, which may be relevant for  intermediate mass UHECR nuclei and nG scale intergalactic magnetic fields with 1 Mpc coherence scale.
We found that the trajectories form a non-trivial caustic-like pattern with strong deviation from isotropy.  { Thus, measurements of the flux from a source at a given distance will depend on the position of the observer.}
\end{abstract}

\maketitle

\section{Introduction}

The propagation of cosmic rays in a turbulent magnetic field in the diffusion regime has been studied in detail \cite{Casse:2001be,Giacinti:2012ar,Giacinti:2017dgt}. However, the propagation of Ultra-High Energy Cosmic Rays (UHECR) in the small-angle scattering regime has attracted less attention.  This regime is usually considered  trivial, and in the absence of a regular field, cosmic rays are expected to simply form a blurred image of the source if it is at a distance much greater than the coherence length $D \gg \lambda_C $. The average distribution of UHECRs around their sources in this case was obtained in Ref.\cite{Harari:2015mal}.

Non-trivial lensing effects during UHECR propagation in the Galactic magnetic field were discovered in Ref.\cite{Harari:2000he,Harari:2002dy}. In particular, in Ref.~\cite{Harari:2002dy} these effects were studied for the case of initially parallel cosmic rays passing through a turbulent magnetic field of the Galaxy.  Similar lensing effects for a magnetic field in a cluster of galaxies were studied in Ref.~\cite{Dolag:2008py}.

{In this paper, we investigate the propagation of UHECRs from their sources in a certain range of parameters of the intergalactic magnetic field (IMF). It is shown that even in the $D\gg\lambda_C$ regime, the outcome of UHECR propagation is not isotropic, and depending on the position of the observer, the source fluxes can significantly deviate from their average values.}

The paper is organized as follows. In Section 2, we discuss average deflections of UHECRs from their original directions. In Section 3, we study variations in the UHECR flux density due to deflections in the turbulent field at a given distance from the source.

\section{Deflection angles of UHECRs in  turbulent IGMF}
\begin{figure}
    \includegraphics[width=\linewidth]{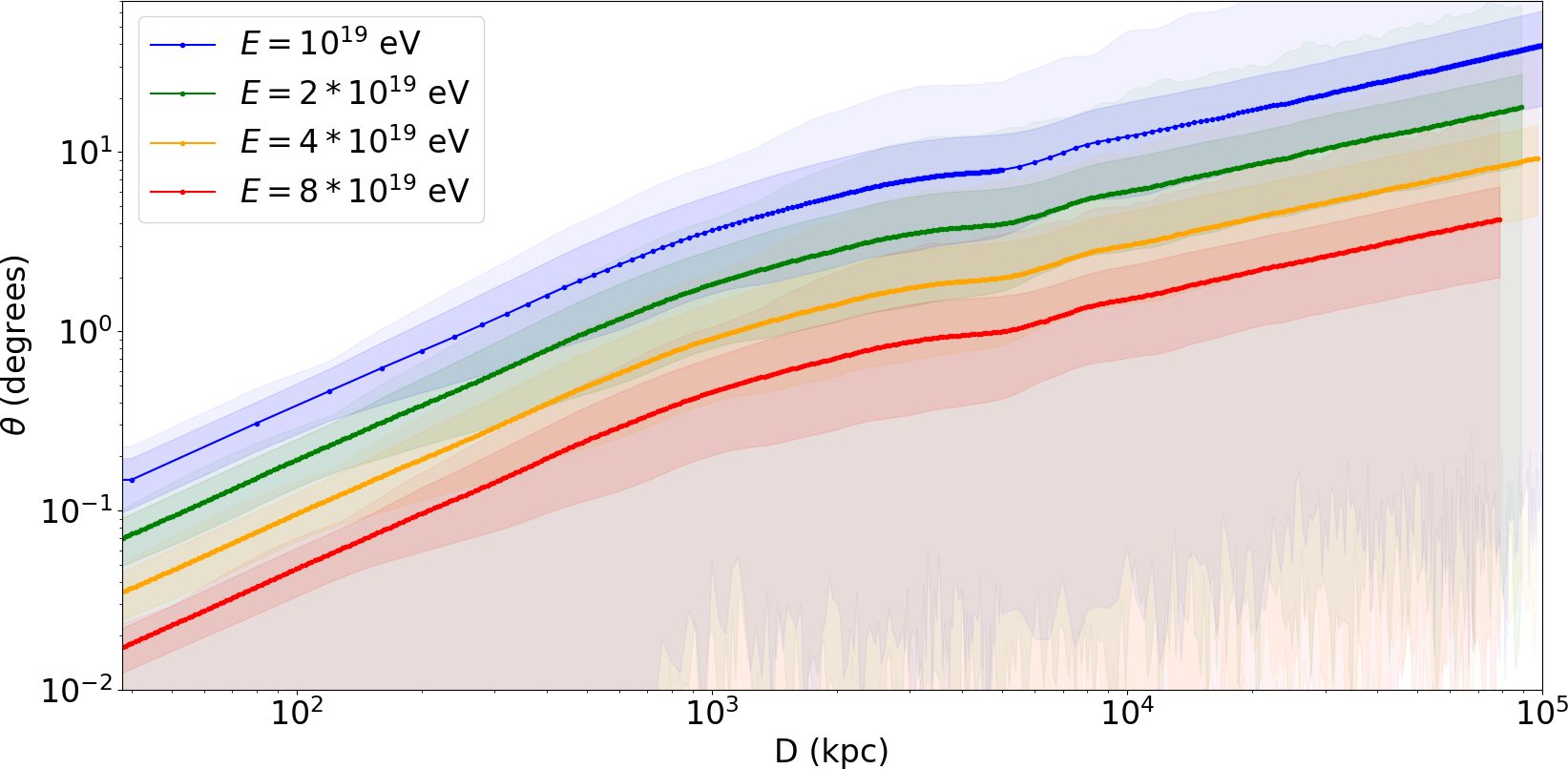}
    \includegraphics[width=\linewidth]{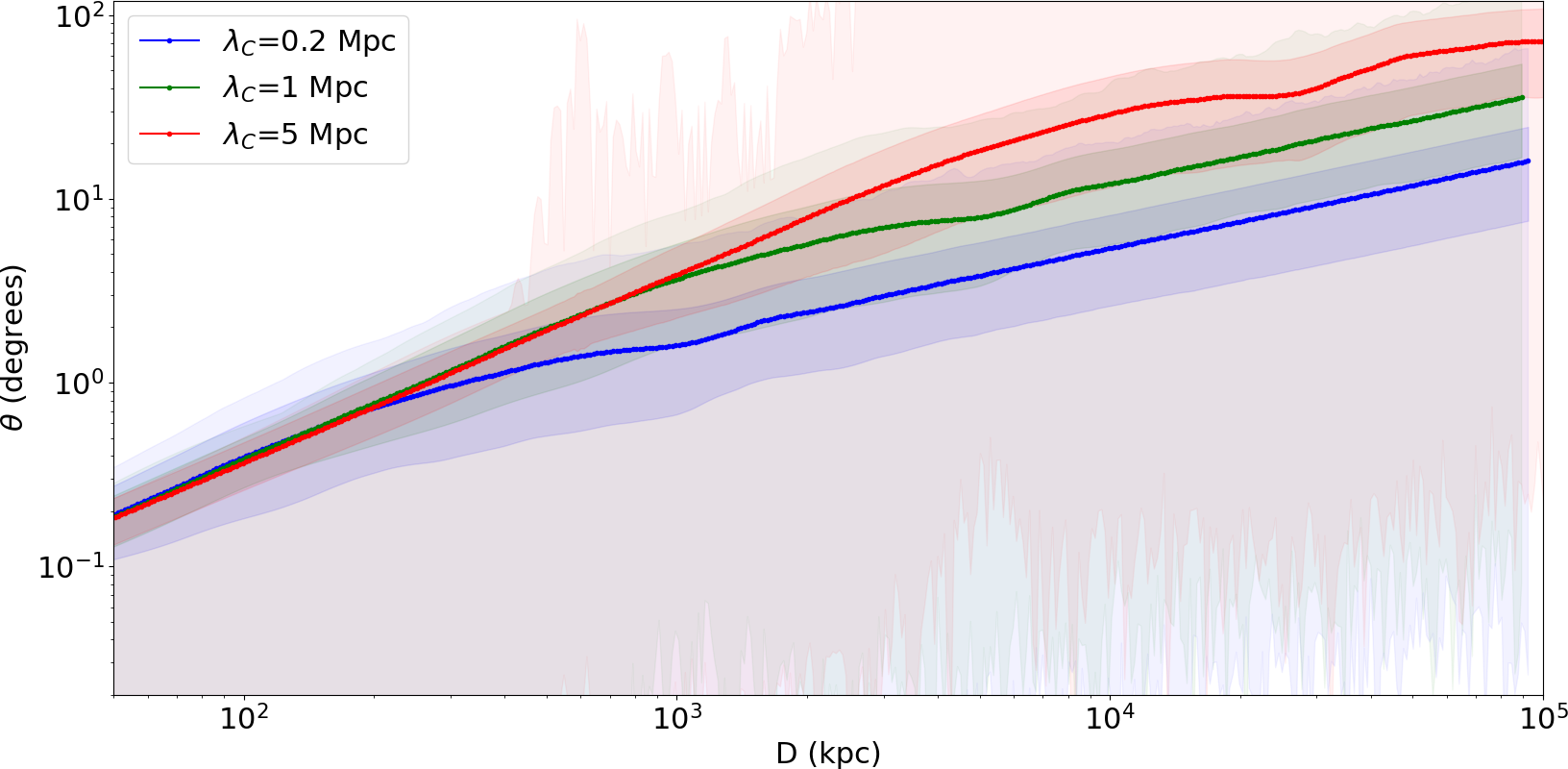}
   \caption{Top panel: Average deflection angle of UHECR protons from the initial direction as a function of distance from the source for $\lambda_C=1$ Mpc and $B=1$ nG and several UHECR energies  $E = 10, 20, 40, 80$ EeV. Semitransparent filling denotes min-max and standard deviations of deflection angles. Bottom panel: Average deflection angle of UHECR protons from the initial direction as a function of distance from the source for $E=10$ EeV and $B=1$ nG and several coherent lengths  $\lambda_C = 0.2, 1, 5$ Mpc.}
    \label{fig:Energy}
\end{figure}

Deflection of cosmic rays in a turbulent field with coherence length $\lambda_C$ as a function of distance has three regimes. At $D<\lambda_C$, the deflection occurs across the main direction of the magnetic field in the given region. As a result, the deflection grows almost linearly $\theta = (D/D_0)^\alpha$ with $\alpha=0.9$ in our case. For $D>5\lambda_C$ deflections occur in diffuse regime and a typical power law corresponds to $\alpha=0.5$. Finally, in the intermediate regime $\lambda_C<D<5\lambda_C$, the exponent $\alpha$ is interpolated between the above values.

\begin{figure}
    \includegraphics[width=\linewidth]{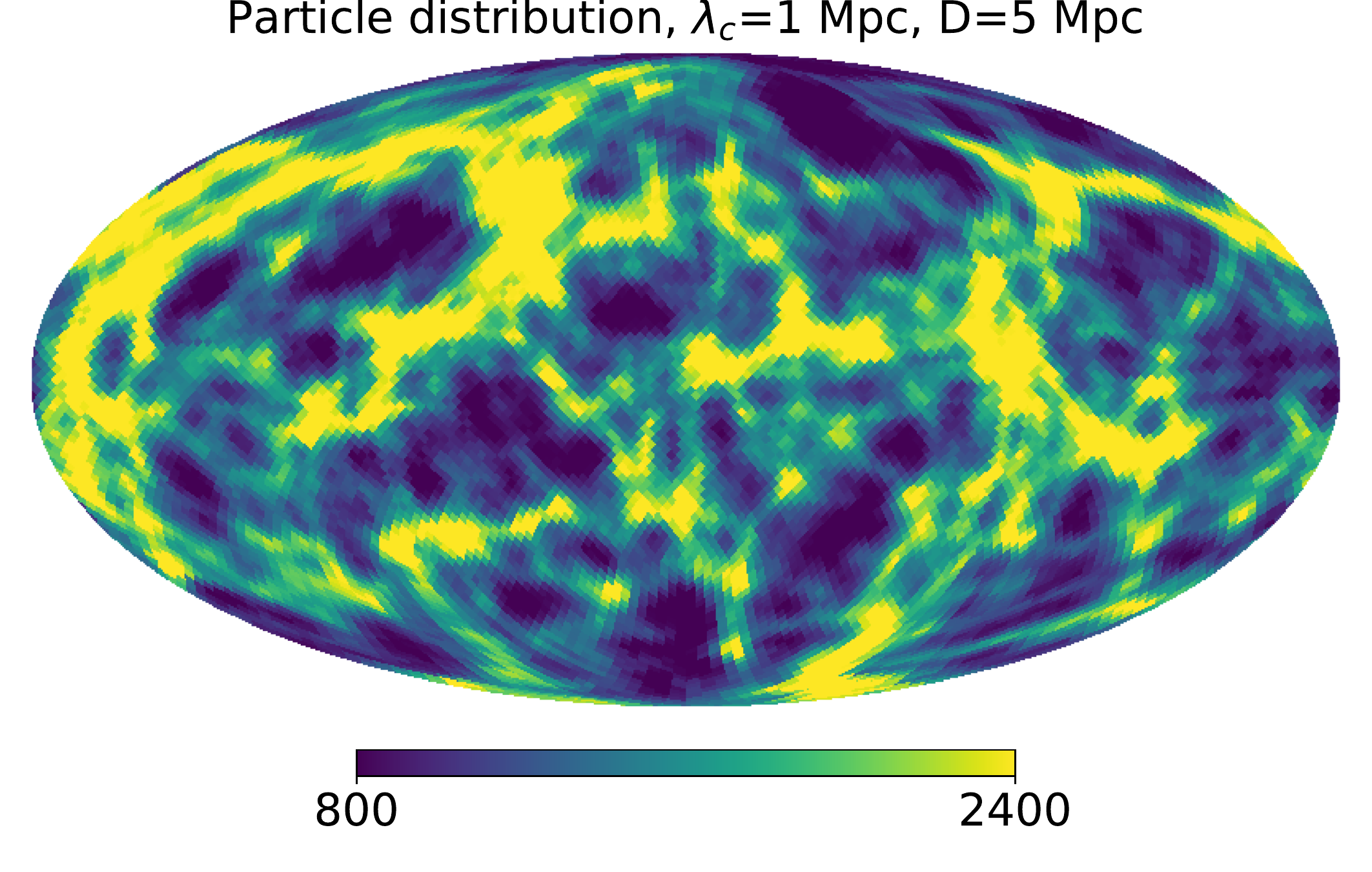}
    \includegraphics[width=\linewidth]{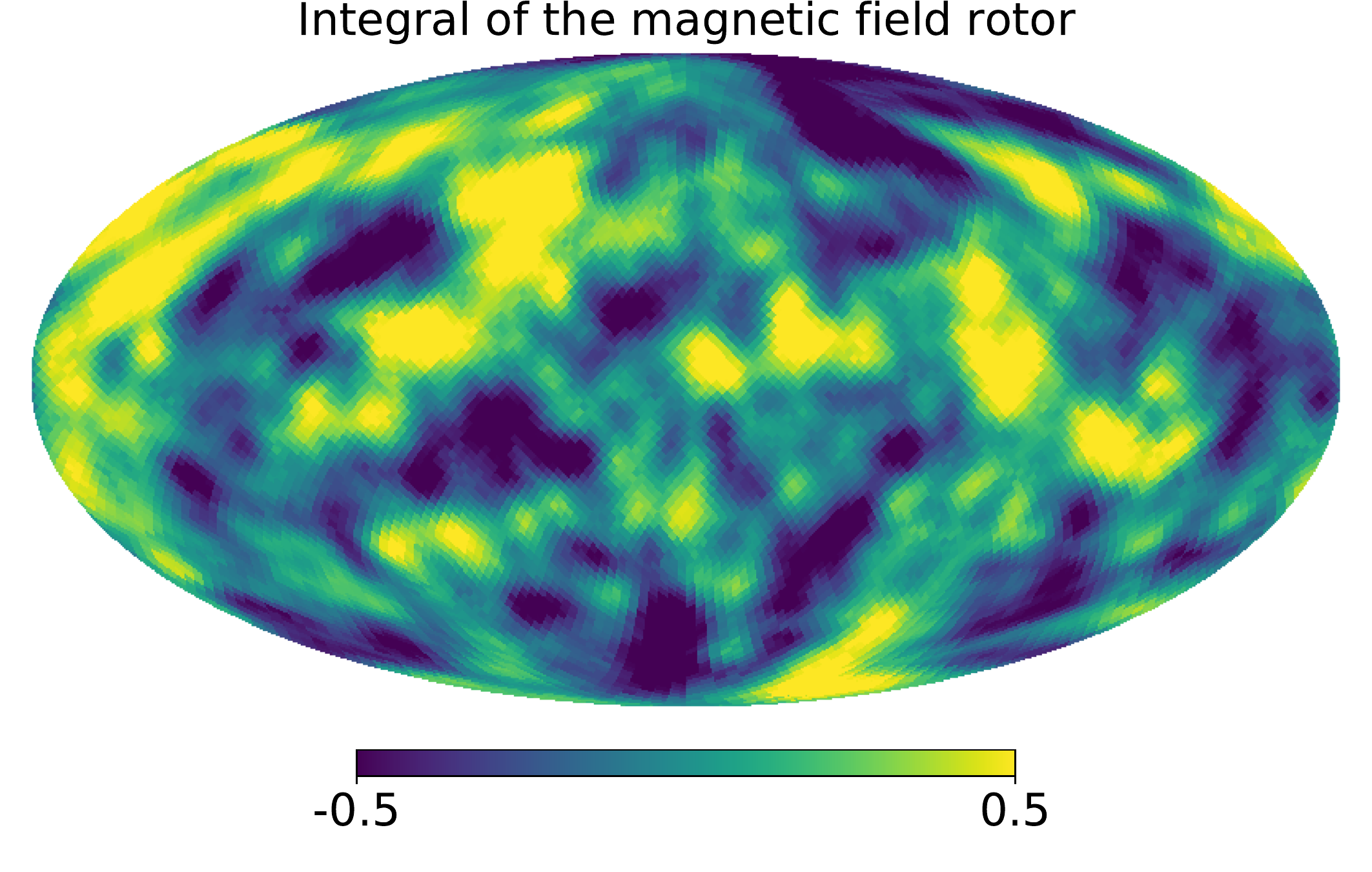}
   \caption{Top panel: distribution of particles on the sphere with radius of 5 Mpc. Bottom panel: Integral of the magnetic field rotor.}
    \label{fig:counts_and_rotB_maps}
\end{figure}
\begin{figure}
    \includegraphics[width=0.958\linewidth]{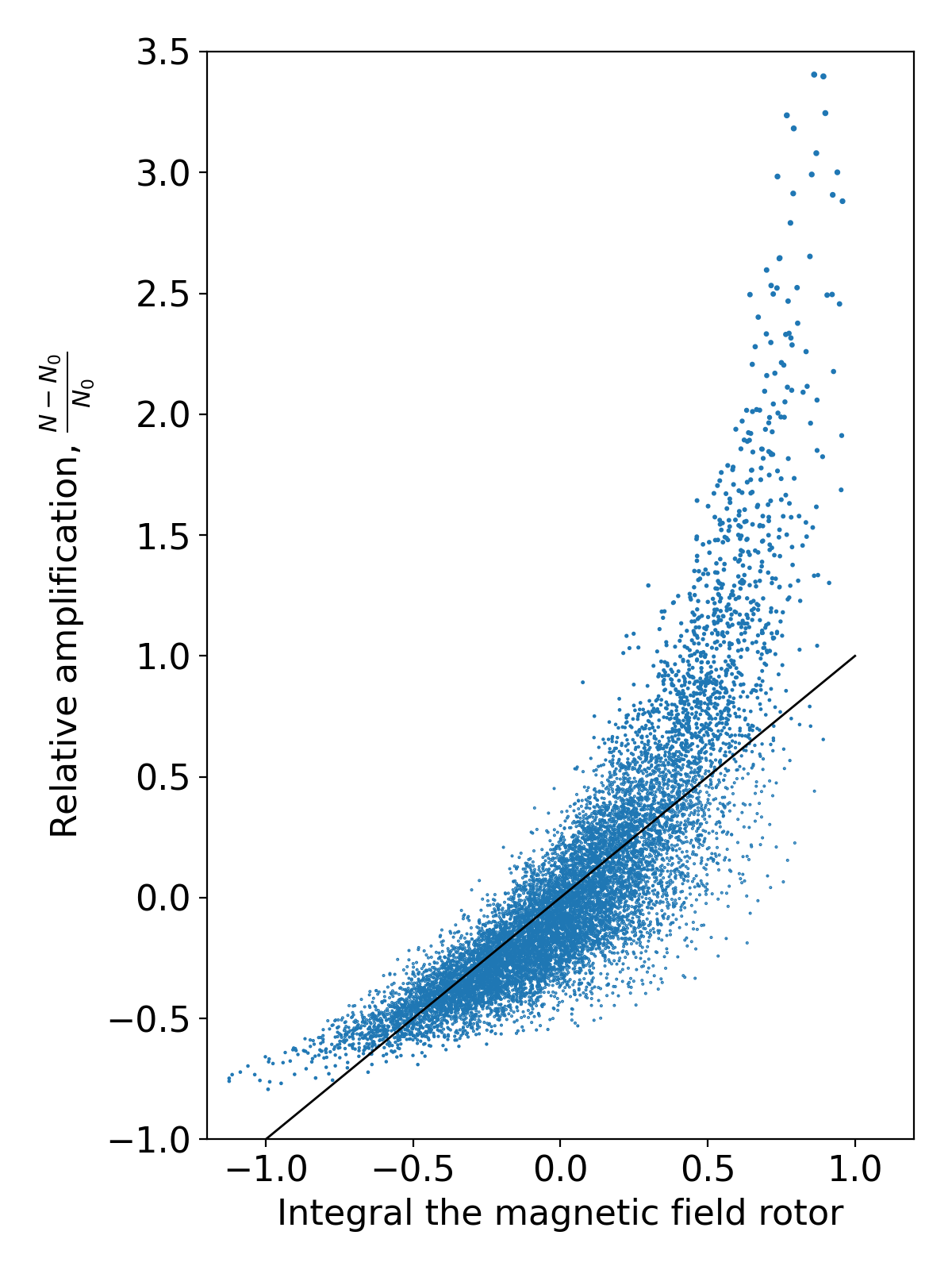}
   \caption{Relative amplification of the number of particles in a pixel as a function of the integral of the rotor of the magnetic field in the direction of that pixel.}
    \label{fig:counts_vs_rotB}
\end{figure}

Therefore, for $D\gg\lambda_C$ and small deflection angles, we numerically expect
\begin{equation}
\theta \sim 4^\circ \; Z \frac{B}{\rm nG}\;  \frac{10 ~ \rm EeV}{E} \sqrt{\frac{D}{\rm Mpc}} \sqrt{\frac{\lambda_C}{\rm Mpc}}
\label{theta_average}
\end{equation}
where $Z$ is the atomic number of the primary particle.

The distance dependence of the proton deflection angle in the other two regimes is shown in Fig.~\ref{fig:Energy}, top panel, for $\lambda_C=1$ Mpc and $B=1$ nG and several UHECR energies. As expected, an increase in energy reduces the scattering angle. The average deflection angle on the sphere is inversely proportional to the particle energy at all distances to the source, including the intermediate regime $D \sim \lambda_C$.

Similarly, the dependence on the coherence length $\lambda_C$ is shown in Fig.~\ref{fig:Energy}, bottom panel, for $E=10$ EeV and $B=1$ nG. Blue, green and red lines show the mean deflection for $\lambda_C=0.2$ Mpc, $\lambda_C=1$ Mpc and $\lambda_C=5$ Mpc respectively. For greater distances $D\gg\lambda_C$, deflections obey the equation~\ref{theta_average}.

Relations for mean deflections of cosmic rays presented in this section are well known, however, averaging can erase important properties of UHECR propagation at a given distance from the  source. In the next section, we discuss the two-dimensional UHECR flux distribution on the sphere around the source and its dependence on the parameters of the turbulent magnetic field.

\section{Flux density of UHECR on the sphere around the source}
The angular distribution of cosmic rays around the original direction from the source, averaged over the azimuthal angle, was studied in Ref.\cite{Harari:2015mal}. The propagation of initially parallel cosmic rays in the Galaxy both in regular and turbulent fields was studied in \cite{Harari:2000he,Harari:2002dy}.

However, two-dimensional slices of the density distribution of cosmic rays propagating from a source have never been studied. In this section, we study the dependence of the two-dimensional distribution of cosmic rays on the sphere around the source as a function of their energy $E$ and distance from the source $D$, and also as a function of the turbulent magnetic field coherence length~$\lambda_C$. Since the field strength $B$ is degenerate with energy, we explicitly study the dependence on  the three parameters indicated above only.

\subsection{Weak amplification}

We start with a simplified model of a source emitting protons isotropically in all directions. The entire volume around the source is filled with the homogeneous turbulent magnetic field. Let us also assume that the parameters of the magnetic field and the energy of the particles are chosen in such a way that the deflections after propagating a distance of several correlation lengths are small, but not negligible. In this model, one would expect that the distribution of UHECR arrival positions on a sphere of radius $D > \lambda_C$ would remain isotropic, since the deflections of particles initially aimed in different directions are statistically independent in a random magnetic field. However, as we  show below, this is true only if the radius of the sphere is much larger than the correlation length. On the other hand, if the radius of the sphere is on the order of several correlation lengths,  the distribution of particles on the sphere becomes essentially anisotropic at intermediate angular scales.

The physical reason for the appearance of inhomogeneities during propagation in a turbulent field can be understood analytically. In \cite{Harari:2002dy} it is shown that in the case of  initially parallel proton beam, inhomogeneities in the particle distribution are caused by fluctuations of the magnetic field rotor along the trajectories of neighboring particles. Here we follow the logic of \cite{Harari:2002dy} and derived the modified formula for the flux amplification for the case of a divergent particle beam.

\begin{figure*}
    \includegraphics[width=\textwidth]{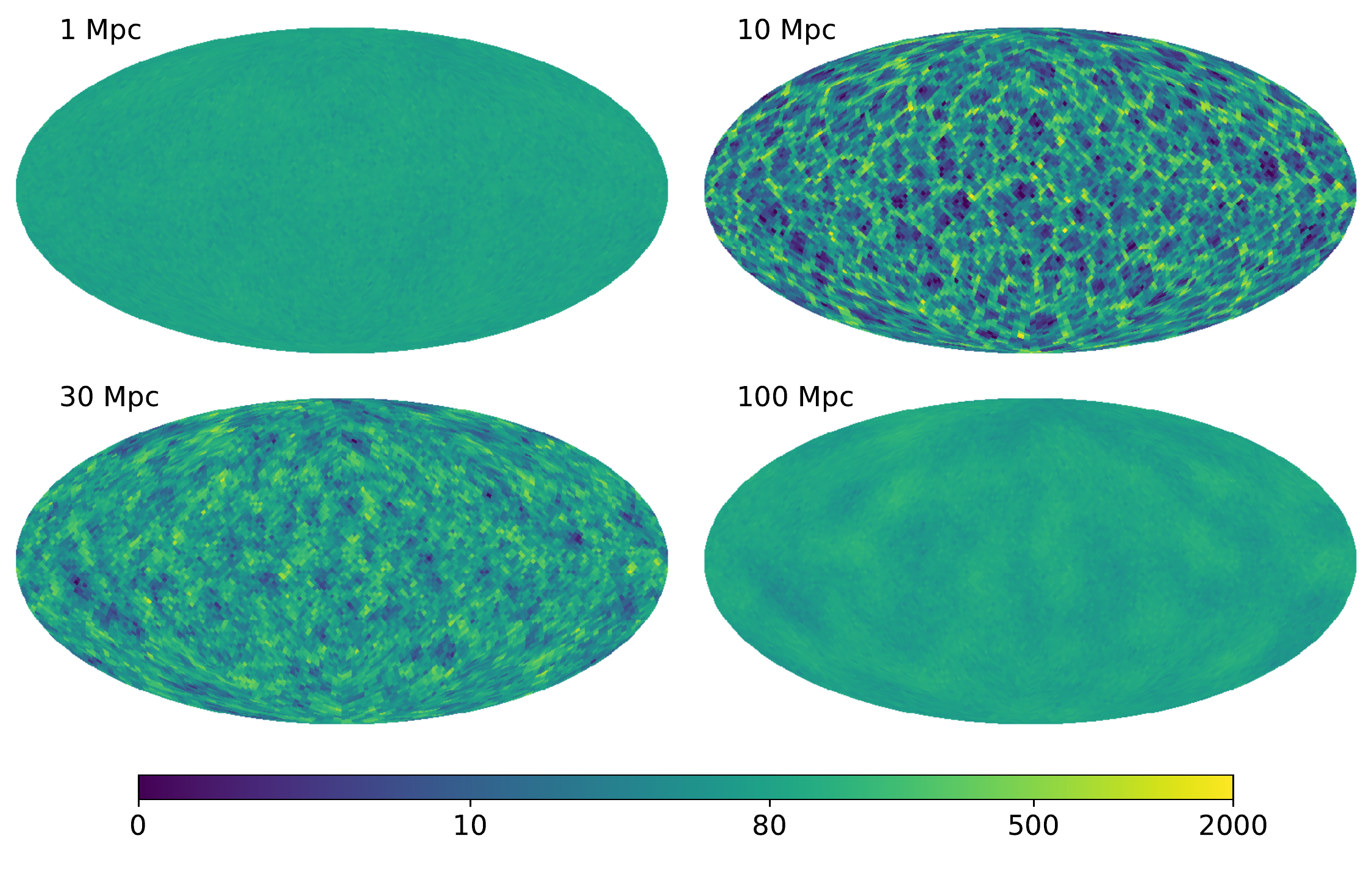}
   \caption{Distribution of cosmic rays over the sphere around the source. It can be seen that the medium-scale anisotropies first increase and then blur. Different panels represent different sphere radii: 1 Mpc, 10 Mpc, 30 Mpc and 100 Mpc, respectively. Magnetic field strength $B=1$ nG, correlation length $\lambda_C=0.3$ Mpc, proton energy $E=10$ EeV. Note that here the sky maps are plotted on a logarithmic scale, and the average number of particles per pixel is 80.}
    \label{fig:maps_together}
\end{figure*}
\begin{figure*}
    \includegraphics[width=\textwidth]{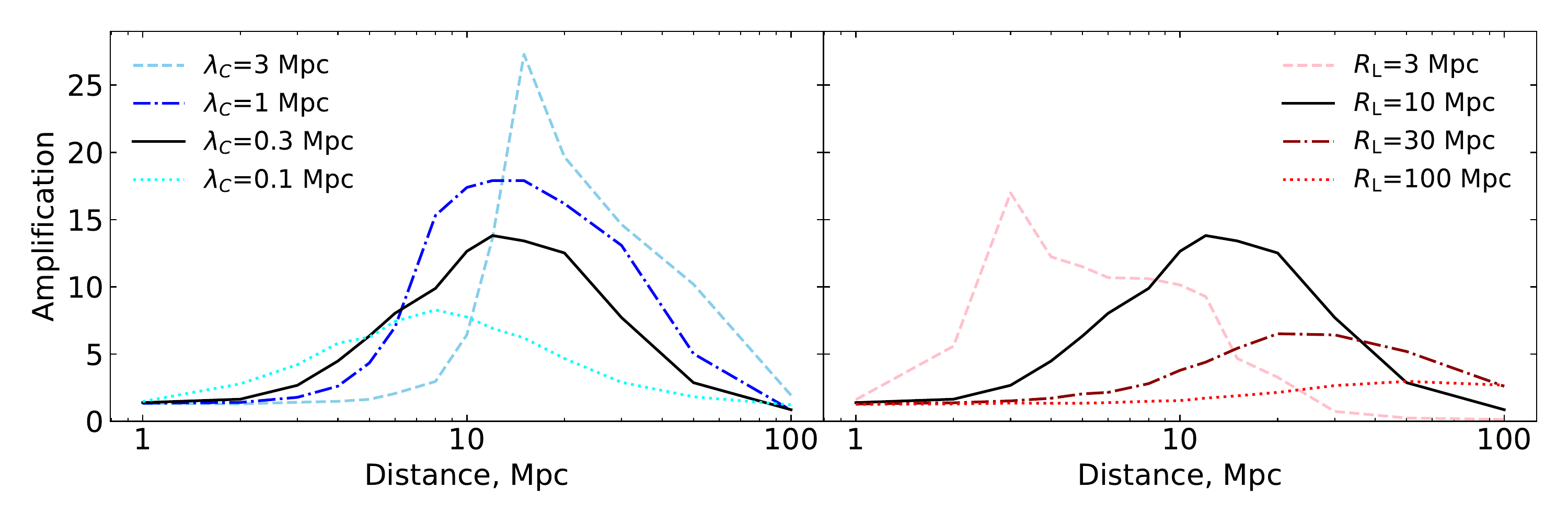}
   \caption{The evolution of the flux with distance, averaged over the five brightest spots on the sphere. Left panel: Flux gain over average flux on a sphere as a function of source distance for a fixed Larmor radius $R_L=10$ Mpc for several coherence lengths $\lambda_C=0.1,0.3,1,3$ Mpc.
   Right panel: the same gain for a fixed coherence length $\lambda_C=0.3$ Mpc for several Larmor radii $R_L=3,10,30,100$ Mpc.}
    \label{fig:foc_dist}
\end{figure*}

Consider two particles shifted by the vectors $\Delta\vec{x}_i$, $i=1,2$ relative to  the reference trajectory $\vec{x_0}$. If the distance between particles is less than the correlation length, the evolution of the displacement vector $\Delta\vec{x}_i$ follows Equation 4.1 from \cite{Harari:1999it}:
\begin{equation}
	\frac{\mathrm{d}^2\Delta \vec{x}_i}{\mathrm{d}t^2} = \frac{Zec}{E}\left(\frac{\mathrm{d}\vec{x}_0}{\mathrm{d}t}\times\Delta\vec{B}_i(\vec{x}_0)+\frac{\mathrm{d}\Delta\vec{x}_i}{\mathrm{d}t}\times\vec{B}(\vec{x}_0) \right)
\end{equation}
which is a direct consequence of the Lorentz equation. Here $Ze$ is the charge of the particle, $c$ is the speed of light and $\Delta\vec{B}_i(\vec{x}_0) \equiv \vec{B}(\vec{x}_0 + \Delta \vec{x}_i) - \vec{B}(\vec{x}_0)$. To calculate the change in the flux, consider two particles displaced relative to the reference in the perpendicular directions $\Delta y$ and $\Delta z$. The change in flux is inversely proportional to the change in area $A$ subtended by these displacement vectors. For an initially parallel beam of particles, the area between the particles turns out to be linearly dependent on the rotor of the magnetic field \cite{Harari:2002dy}:
\begin{equation}\label{eq:harari}
	A(D) = A_0 \left(1 - \frac{Ze}{E}\int\limits_0^D (D-s)\,(\mathrm{rot}\vec{B}\cdot\mathrm{d}\vec{s}) \right)
\end{equation}
where $s$ is a parameter along the trajectory.

In the case of a diverging beam, the equation (\ref{eq:harari}) should be modified due to different beam geometry, which determines different initial conditions. We have used these initial conditions and have derived the equation for the diverging beam:
\begin{equation}\label{eq:new_harari}
	A(D) = A_0 \left(1 - \frac{Ze}{E}\int\limits_0^D s\left(1-\frac{s}{D}\right) (\mathrm{rot}\vec{B}\cdot\mathrm{d}\vec{s})\right)
\end{equation}
It should be noted that the equations are valid only for small amplification $(A(D)-A_0)/A_0 \ll 1$ and small deflections of particles.

\subsection{Numerical simulations}
Equation \ref{eq:new_harari} provides a direct prediction for the flux amplification that can be tested numerically. Indeed, if $N_0$ is the mean number of particles that have passed through a given pixel, then the number of particles $N$ that actually hit a given pixel corresponds to the amplification of the flux. Rewriting equation \ref{eq:new_harari} in terms of the number of particles, we  arrive to a relation that can be verified directly in numerical simulations:
\begin{equation}
    \frac{N - N_0}{N_0} = \frac{1}{R_L} \left[\frac{\int\limits_0^D s\left(1-\frac{s}{D}\right) (\mathrm{rot}\vec{B}\cdot\mathrm{d}\vec{s})}{B}\right]\,,
\end{equation}
where $R_L$ is a Larmor radius.

We use publicly available codes \texttt{CRbeam} \cite{Berezinsky:2016feh} and \texttt{CRPropa} \cite{AlvesBatista:2016vpy,AlvesBatista:2022vem} for cosmic ray propagation. All interactions and cosmological expansion of the Universe are turned off. The particles were emitted isotropically and propagated until they reached a sphere of a given radius~$D$. For each particle, we store its initial direction, final direction, and final position on the sphere. The resulting output is processed with the package \texttt{healpy}~\cite{Zonca2019,2005ApJ...622..759G}.

The use of two codes makes it possible to model a turbulent magnetic field in two different ways. In \texttt{CRbeam} the magnetic field is generated as a sum of plane waves with random phases and directions \cite{1999ApJ...520..204G} and its exact value is recalculated on the fly before each next step during particle propagation. In \texttt{CRPropa}, on the contrary, the field is precalculated on the grid and its value is set by interpolation between grid points. In both codes, we generate a magnetic field with a Kolmogorov spectrum and a minimum scale 100 times smaller than the maximum scale.

In all our simulations, \texttt{CRbeam} and \texttt{CRPropa} produced consistent results. Given the fact that the generation of magnetic fields and the propagation of particles in these codes are carried out in completely different ways, this increases the reliability of the results.

For the first test, we set the magnetic field strength to $B=1$~nG, $\lambda_C=1$~Mpc and the proton energy to $E=10$~EeV. The radius of the final sphere has been chosen to be $D=5$ Mpc. The results are shown in the Figures \ref{fig:counts_and_rotB_maps} and \ref{fig:counts_vs_rotB}. Upper panel of the Figure \ref{fig:counts_and_rotB_maps} shows the number of particles in each pixel of the \texttt{healpy} skymap. The average number of particles $N_0$ expected in each pixel is 1600 and the color scale is adjusted to the range of 800 - 2400 particles, which corresponds to a relative deviation of 0.5 from the mean. The lower panel shows the calculation of the integral of the rotor of the magnetic field, i.e. the pixel value corresponds to the r.h.s. of equation \ref{eq:new_harari}, calculated along the radius of the sphere directed towards the center of the given pixel. The color scale is also adjusted to the range -0.5 - 0.5, allowing one-to-one comparison of the graphs. It can be seen that the positions of particle density fluctuations coincide with the fluctuations of the integral of r.h.s of equation \ref{eq:new_harari}.
\begin{figure}
    \includegraphics[width=1\linewidth]{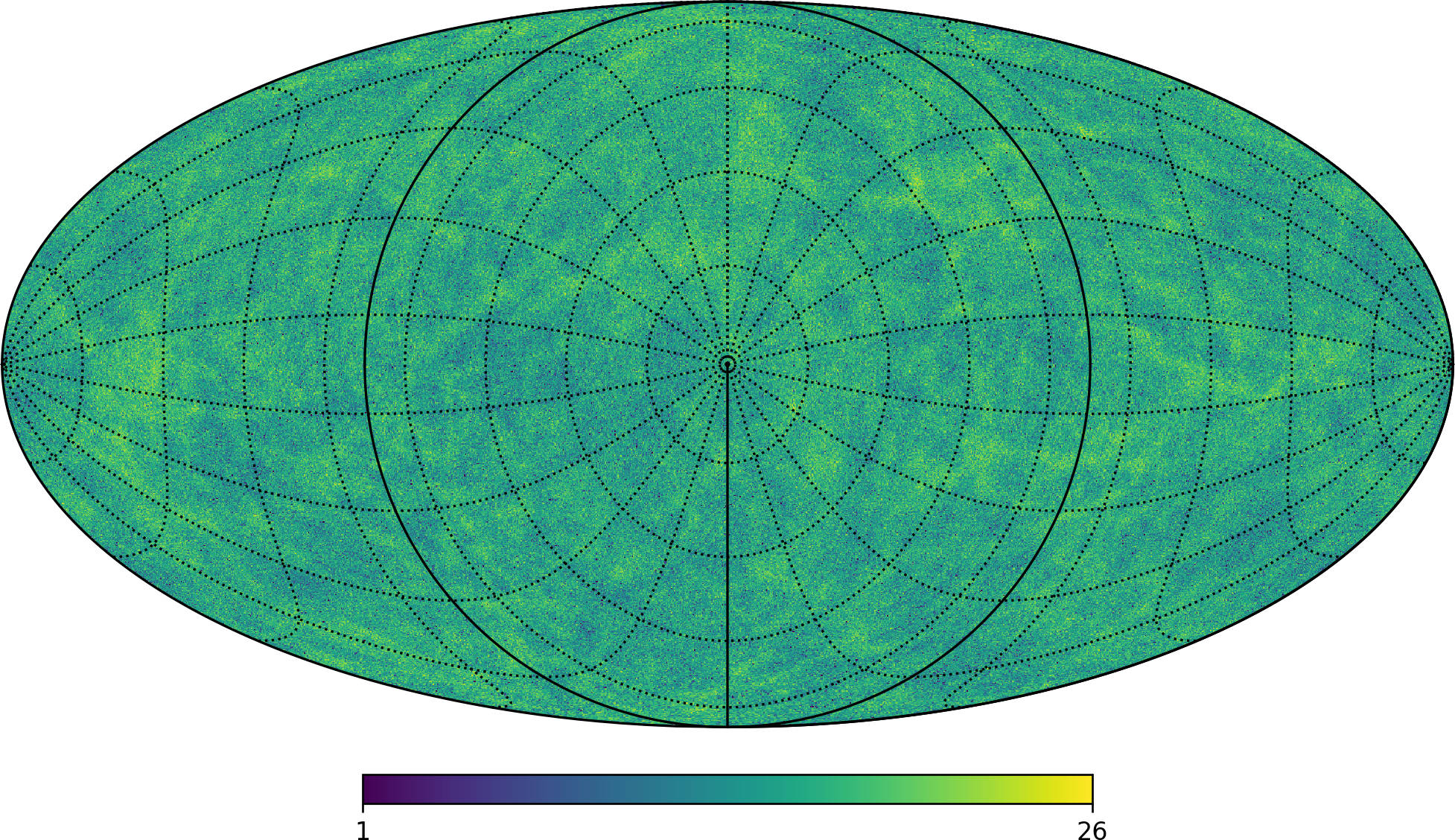}\hfill
    \\[\smallskipamount]
    \includegraphics[width=1\linewidth]{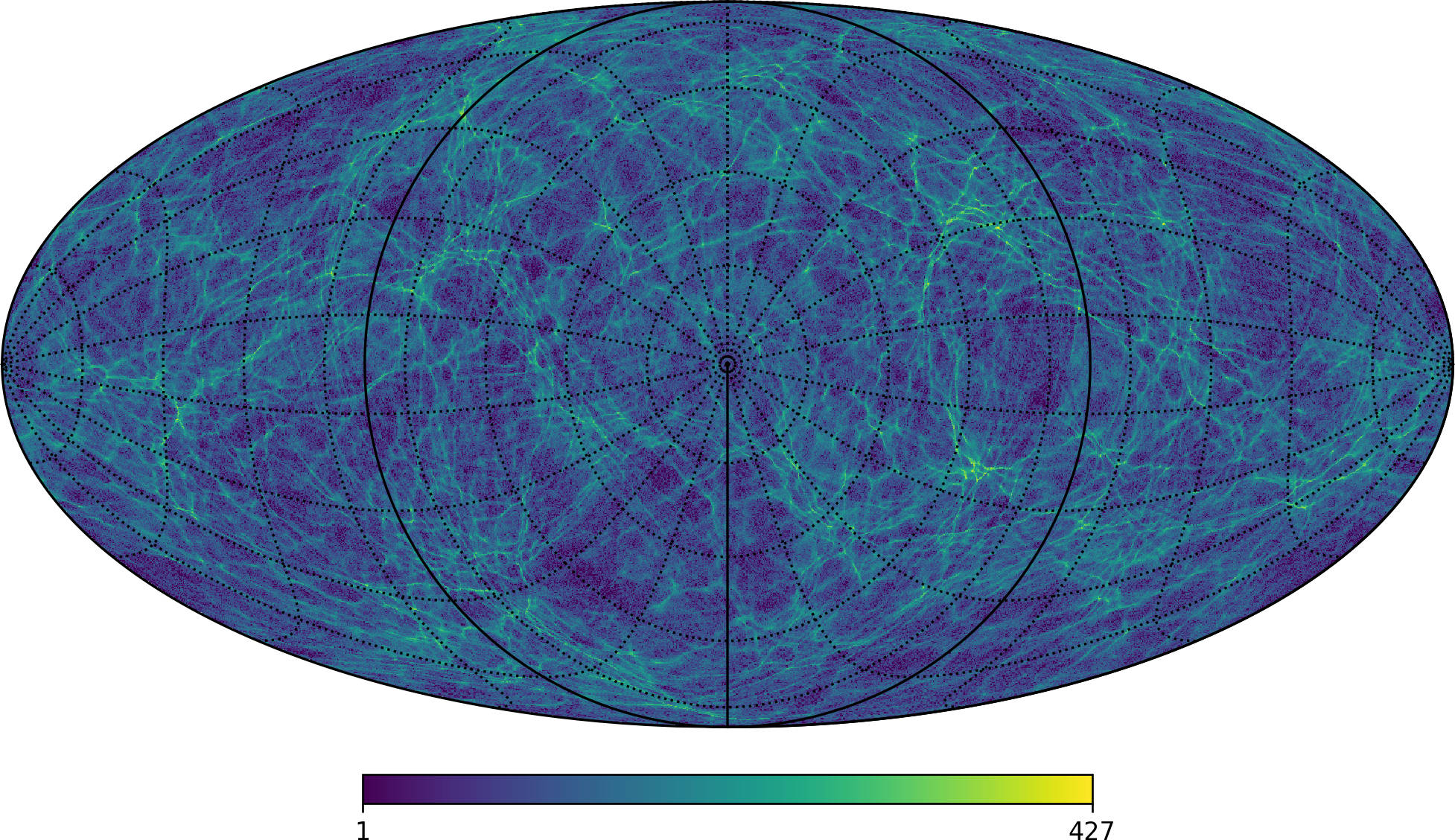}\hfill
    \includegraphics[width=1\linewidth]{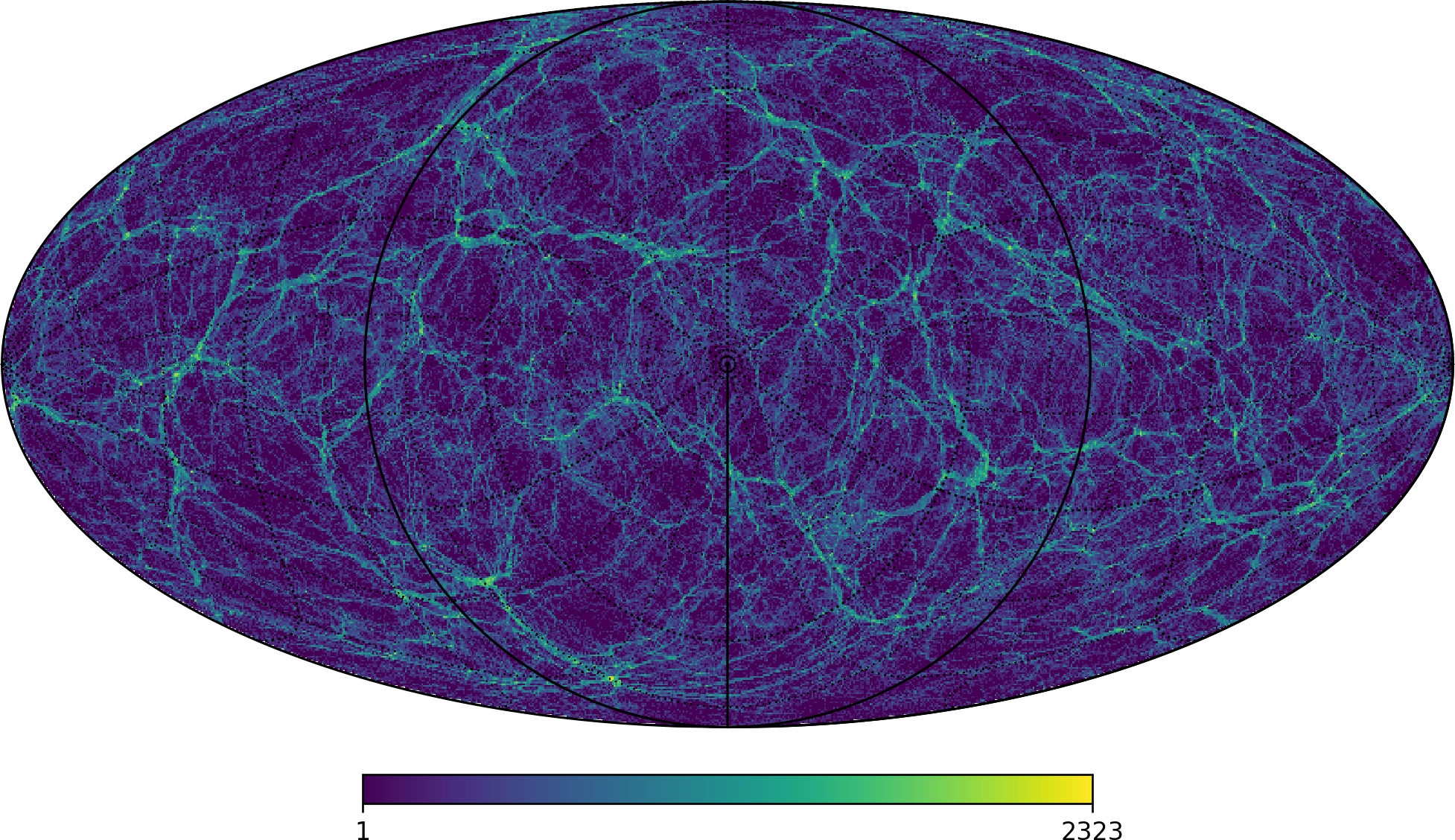}\hfill
    \caption{UHECR distribution on a sphere with a radius of 10~Mpc. The magnetic field was turned off after 1,3,10 correlation lengths for panels from top to bottom, respectively. We see that the global structure is created after the passage of the first 2-3 correlation lengths, then it simply sharpens and intensifies.}\label{fig:skymap}
\end{figure}
\begin{figure}
    \includegraphics[width=\linewidth]{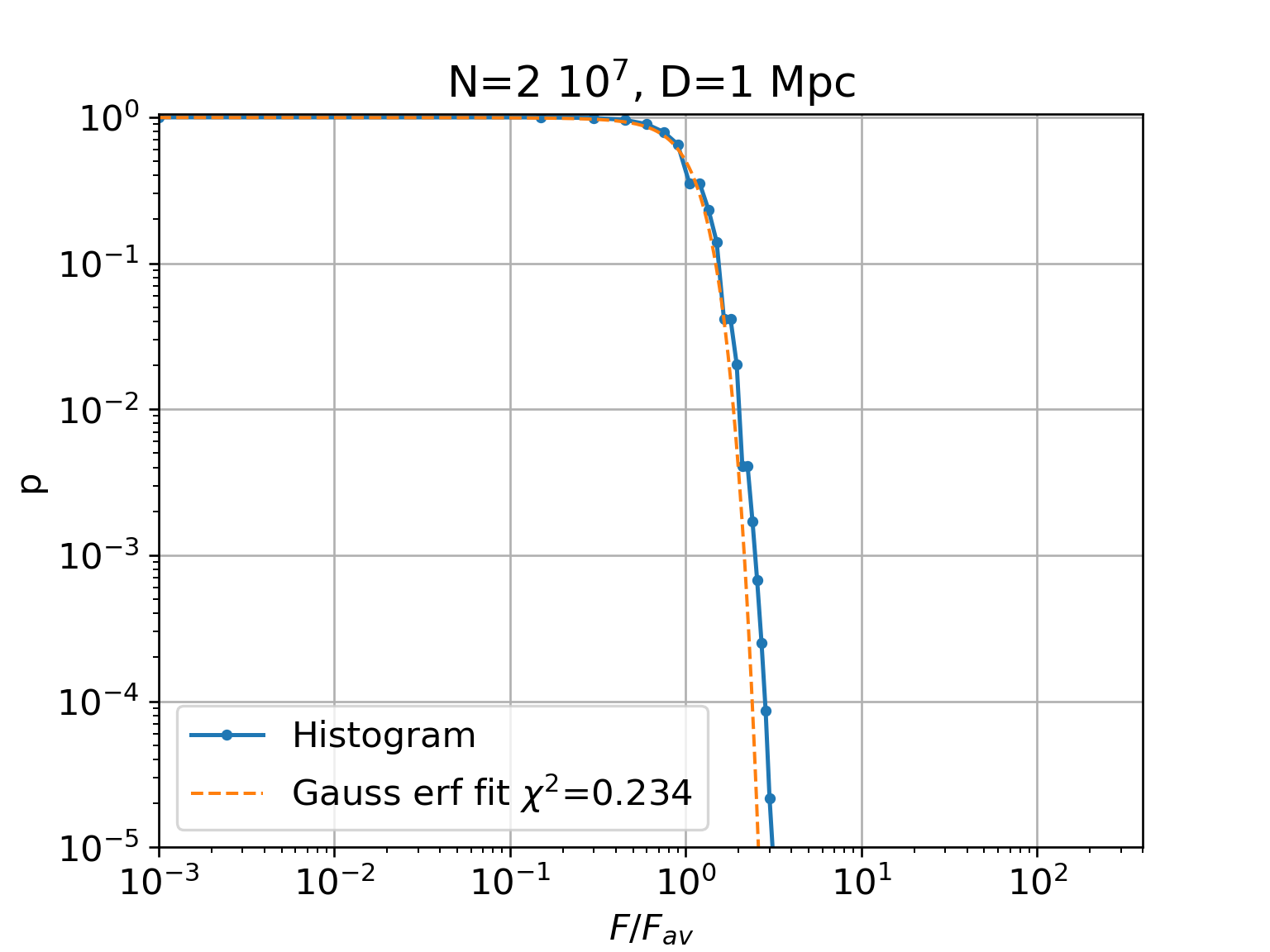}
    \includegraphics[width=\linewidth]{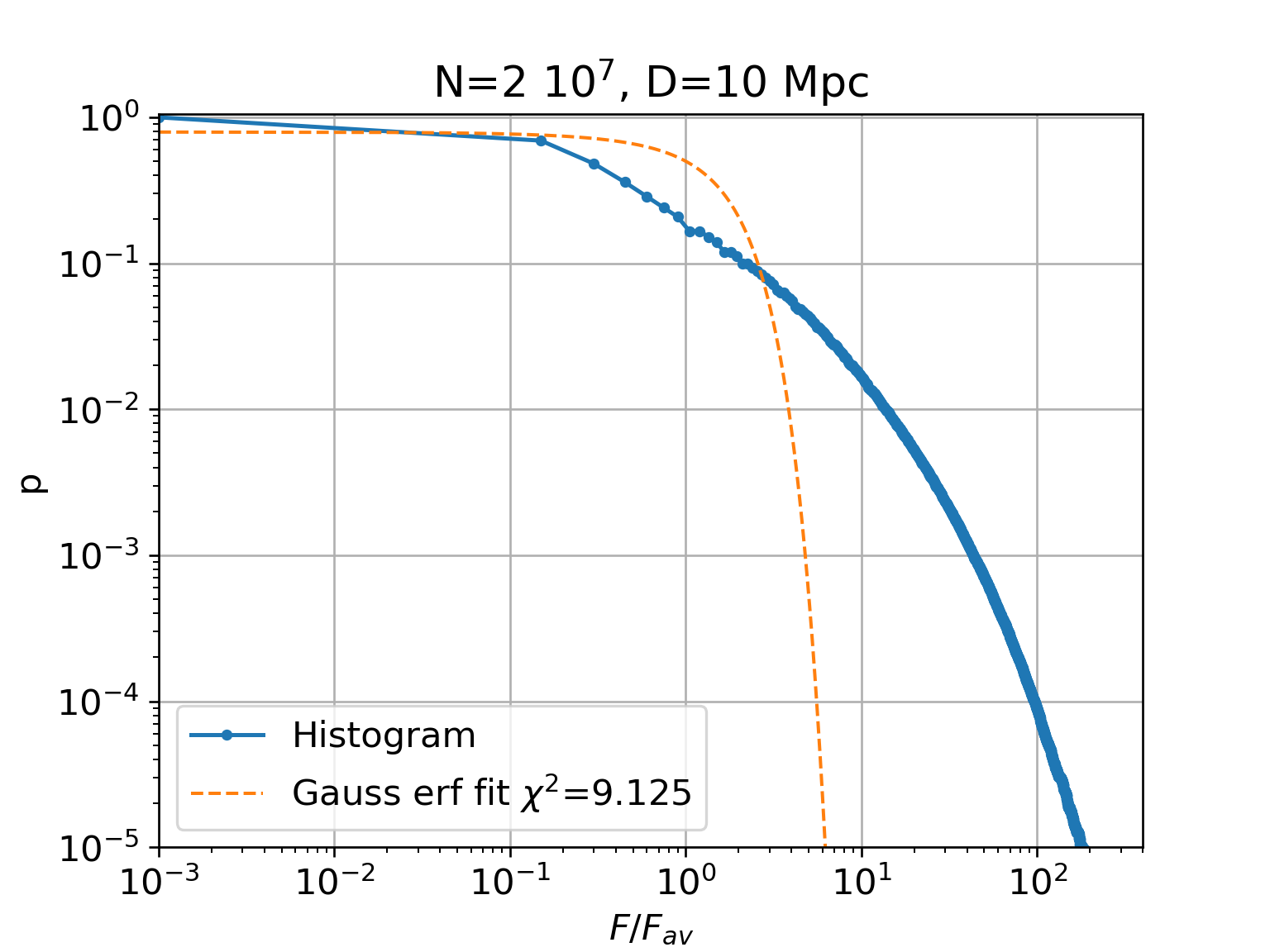}
    \includegraphics[width=\linewidth]{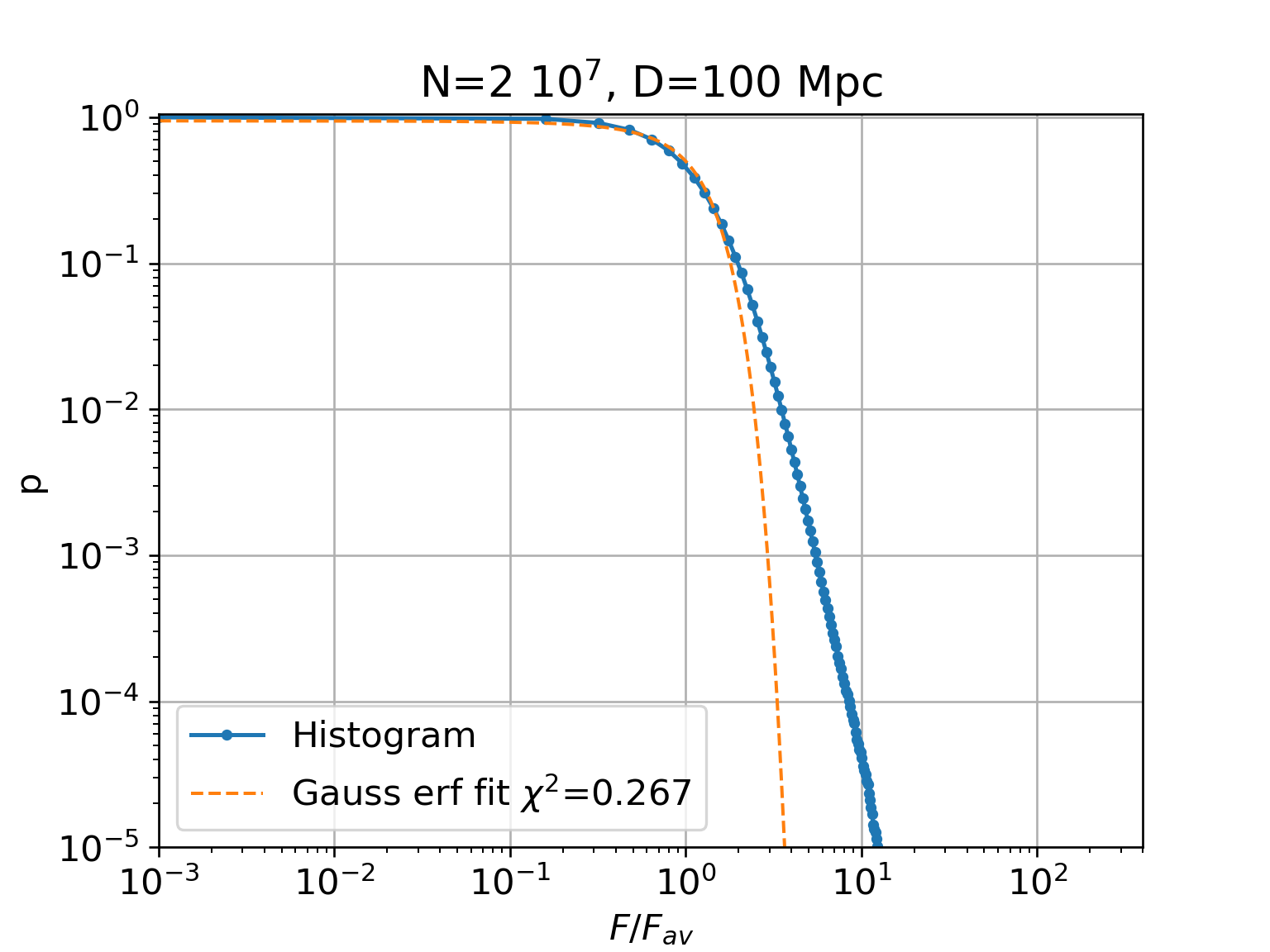}
   \caption{The cumulative probability of arrival with a flux above a given value in units of the average flux on the sphere for an observer located at a distance of 1 Mpc (top panel), 10 Mpc (middle panel) and 100 Mpc (bottom panel) from the source. At the distance close to Larmor radius the distribution is strongly non-Gaussian.}
    \label{fig:distr}
\end{figure}

This is also clearly seen in  Fig.\ref{fig:counts_vs_rotB} where the relative amplification of the number of particles in a pixel is shown as a function of the integral \ref{eq:new_harari} evaluated in the direction to that pixel. One dot on the graph represents one \texttt{healpy} pixel, and the black straight line corresponds to the theoretical prediction of the equation \ref{eq:new_harari}. In the vicinity of zero, the dots follow a theoretical linear law, however, in the region of large values of the integral, the gain becomes much stronger.

 In Fig. \ref{fig:maps_together} the evolution of medium-scale anisotropies of the distribution of cosmic rays on a sphere is presented as a function of distance to the source for Larmor radius  $R_L = 10$ Mpc (for $B=1$ nG and $E = 10$ EeV) and coherence length $\lambda_C =0.3$ Mpc. The top left panel shows an almost isotropic distribution for $D=1$ Mpc from the source. At 10 Mpc, top right panel, anisotropy is strongest and sharpest, then fades away by 100 Mpc, see bottom panels.

In Fig. \ref{fig:foc_dist} we show how the flux averaged over  the five brightest spots  on the sphere changes as a function of the distance to the source for changing values of two parameters: Larmor radius $R_L$ and coherence length  $\lambda_C$. In the left panel, we fix the Larmor radius equal to $R_L=10$ Mpc and study the dependence of the gain on distance for several coherence lengths $\lambda_C =0.1,0.3,1.3$ Mpc. It can be seen that with an increase in the coherence length, the gain maximum is reached at a greater distance from the source, but  varies within 0.8 and 1.5 $R_L$. For the right panel of Fig. \ref{fig:foc_dist} the coherence length $\lambda_C =0.3$ Mpc is fixed and the distance dependence of the gain is studied for several Larmor radii $R_L=3,10,30,100$ Mpc.
In all cases, the flux enhancement as a function of distance rises to a maximum at a distance from the source of about 1-2 Larmor radius $R_L$, and then drops to an average value after about 10 $R_L$.

In Fig. \ref{fig:skymap}  we show a sky map with UHECR density at a fixed radius of the sphere around the source. This figure shows an example of a UHECR with $E=10$ EeV propagating in a turbulent magnetic  field with strength $B=1$ nG and coherence scale $\lambda_C = 1$ Mpc. With such a field and energy of cosmic rays, we expect that the maximum anisotropy in the UHECR distribution will be reached at 10 correlation lengths, i.e. at a distance of $D=10$ Mpc. This distance to the source corresponds to the lower panel. The bar under the figure shows the color correspondence to the density of cosmic rays in the structures. Three types of structures can be identified: knots with the highest density, filaments and voids with the lowest density. Note that the typical size of a magnetic field domain with $\lambda_C = 1$ Mpc has an angular scale of 6-10 degrees on this map and corresponds to small features. However, the most prominent and highly visible are the medium-scale structures, with a typical scale of about 60 degrees. To understand the angular scale of these structures, we have shown in Fig. \ref{fig:skymap} the UHECR density at $D=10$ Mpc, but when the magnetic field is set to zero outside the 1 and 3 Mpc spheres, see top and middle panels respectively. It can be seen that most of the global structure observed at a distance of 10 Mpc is formed during the passage of the first 3 correlation lengths from the source, its physical size corresponded to the correlation length at that time. Later on the structure mainly sharpens.

It is clear that the observer stands only at one given point on the sphere. By randomly changing its location, we calculate the cumulative probability of observing a flux with an intensity higher than a given value, which is shown in the Fig.~\ref{fig:distr}. The x-axis is normalized to the mean density on the sphere.  We plot the cumulative probability at a distance of 1, 10, and 100 Mpc from the source, from the top panel to the bottom, respectively for $\lambda_C=1$~Mpc, $B=1$~nG, and $E = 10$~EeV.
Both at small and large distances, the probability distribution is similar to Gaussian. However, in the middle panel it is very far from Gaussian. The probability of getting a flux below average is 80\%. At this distance from the source, the initial flux is likely to decrease, e.g.  by a factor of 10 with a probability of about 20\%. The probability  of  an order of magnitude  gain is only 2\% .

Finally, we have verified that the appearance of caustic-like structures does not depend on specific parameters of the magnetic field spectrum. We first checked that our results are robust to changes in the ratio of maximum and minimum turbulence scales. We reduced the size of the smallest eddy by an order of magnitude and made sure that the results did not change. Secondly, we replaced the turbulent magnetic field with the Kolmogorov spectrum by a field consisting of cells with a size equal to the correlation length. In each cell, the field is uniform with strength $B$, but the direction changes randomly between cells. Repeating the simulation in such a magnetic field, we saw that caustics appear in this case as well.

\section{Conclusions}

In this work, we have studied the propagation of UHECRs in a turbulent intergalactic  magnetic field in the small-angle scattering regime. We found that even if UHECRs are emitted isotropically from their source, they are distributed anisotropically at a distance of the order of the Larmor radius, and again isotropically at a distance 10 times greater. The enhanced regions merge into a filamentary, caustic-like structure on the sphere. The angular arrangement of these regions is dictated by the structure of the magnetic field at several coherence lengths from the source.

For low gains, the distribution of cosmic rays on the sphere can be described analytically by Eq.~(\ref{eq:new_harari}). As can be seen in Fig.~\ref{fig:counts_and_rotB_maps}, this equation describes well the amplification of cosmic rays in the linear regime, but is not suitable for high amplifications. The relative size of the enhanced regions depends on the coherence length of the magnetic field. This anisotropic distribution can affect the UHECR spectrum at distances from the source comparable to the Larmor radius.  In addition, this can also lead to the formation of hot spots in the observer's distribution of cosmic rays over the sky, and can also affect the ratio between the number of observed hot spots and the density of UHECR sources.

\paragraph{Acknowledgements.}  Work of K.D., A.K., G.R. and I.T. was supported by the Russian Science Foundation grant 20-42-09010.  The work of D.S. has been supported  by the French National Research Agency (ANR) grant ANR-19-CE31-0020. Some of the results in this paper have been derived using the healpy and HEALPix packages.



\bibliography{refs.bib}
\end{document}